\documentclass{article}

\usepackage{arxiv}

\usepackage[utf8]{inputenc} 
\usepackage[T1]{fontenc}    
\usepackage{hyperref}       
\usepackage{url}            
\usepackage{booktabs}       
\usepackage{amsmath}
\usepackage{amssymb}
\usepackage{amsfonts}       
\usepackage{nicefrac}       
\usepackage{microtype}      
\usepackage{lipsum}
\usepackage{fancyhdr}       
\usepackage{graphicx}       
\graphicspath{{media/}}     
\usepackage{verbatim}
\usepackage{rotating}
\usepackage{subfig}

\usepackage{natbib}
\setcitestyle{authoryear,open={((},close={))}} 

\usepackage{array}
\newcolumntype{H}{>{\setbox0=\hbox\bgroup}c<{\egroup}@{}}
\title{Reactive Global Minimum Variance Portfolios with $k-$BAHC covariance cleaning}

\author{
  Christian Bongiorno,  Damien Challet \\
Université Paris-Saclay\\CentraleSupélec\\
Laboratoire de Mathématiques et Informatique pour la Complexité et les Systèmes\\
91192 Gif-sur-Yvette, France\\
 \texttt{christian.bongiorno@centralesupelec.fr}\\
}

\begin{document}
\maketitle

\begin{abstract}
We introduce a covariance cleaning method which works well in the very high dimensional regime, i.e., when there are many more assets than data points per asset. This opens the way to unconditional reactive portfolio optimization when there are not enough points to calibrate dynamical conditional covariance models, which happens for example when new assets appear in a market. The method is a $k$-fold boosted version of the Bootstrapped Average Hierarchical Clustering cleaning procedure for correlation and covariance matrices. We  apply this method to global minimum variance portfolios  and find that $k$ should increase with the calibration window length. We compare the performance of $k-$BAHC with other state-of-the-art covariance cleaning methods, including  dynamical conditional covariance (DCC) with non-linear shrinkage. Generally, we find that our method yields better Sharpe ratios after transaction costs than competing unconditional  covariance filtering methods, despite requiring a larger turnover Finally,  $k-$BAHC yields better Global Minimum Variance portfolios with long-short positions than DCC in a non-stationary investment universe.

\end{abstract}


\section{Introduction}

Portfolio optimization requires trustworthy estimation of covariance matrices, or equivalently, of correlation matrices and asset price volatility.  In a dynamical investment context, several problems arise.  First, because the dependence between asset prices is not constant in financial markets, correlations matrices are not constant either, and thus estimation should be dynamical. Investment universes evolve as well and thus it may be worth including in a newly-introduced asset in one's portfolio. A naive idea consists in using as few data points as possible, but this gives rise to very large estimation noise of covariance matrices when the number of data points $\tau$ is comparable to the number of assets $n$ (the so-called curse of dimensionality); even worse, correlation and covariance matrices are non-invertible when $n<\tau$. This explains the need to use an efficient filtering method to reduce one-shot estimation noise of covariance or correlation matrices and to regularize them (see \cite{bun2017cleaning} for a review and the literature overview section below). A more palatable econometric approach is provided by the dynamical conditional correlation (DCC) approach~\citep{engle2002dynamic} and its corrected version \citep{aielli2013dynamic}, which adds a dynamical component to correlation estimation, and its latest variation which includes the filtering of the target matrix \citep{engle2019largecov}.

The second ingredient to improve covariance matrix estimation is to account for stochastic volatility itself by using dynamical methods such as GARCH and its numerous variants. Combining both correlation filtering and conditional volatility estimation with DCC was introduced recently  \citep{engle2019largecov} and is considered state of the art.

Our contribution is to improve the BAHC correlation matrix filtering method \citep{bongiorno2021covariance} and to test it against other filtering methods such as non-linear shrinkage (NLS) and other types of Rotationally Invariant Estimators. Remarkably, our method works well even when $\tau\ll n$ (for example with $\tau=30$ data points and $n=500$ assets) as it produces invertible matrices. The first aim of this paper is to establish how and when the improved BAHC unconditional covariance cleaning method performs better than any other competitive unconditional method. We then compare DCC+NLS with  the new BAHC in non-stationary investment universes.

\section{Literature overview}

We focus here on the global minimum variance portfolio optimization problem (GMV) because it provides the best test of the covariance forecast abilities of each filtering method, as indeed the optimal asset weights only depend on the inverse of the estimated covariance matrix. GMV itself is a special case of minimum variance portfolios, themselves a subset of relevant risk/reward cost functions \citep{markowitz1959portfolio,black1990asset,duffie1997overview,hull1998value,krokhmal2002portfolio,roncalli2013introduction,meucci2015risk}.  

The necessity to filter covariance (and correlation) matrices was recognized a long time ago in this context  \citep{michaud1989markowitz}. This is due for example to the fact that mean-variance optimization places more weight on the smallest eigenvalues of the covariance matrix, which, all other things being equal, are more likely to be purely caused  by estimation noise (see e.g. the discussion in \cite{potters2005financial} and subsection \ref{sec:spec_prop}). According to the Mar\v cenko-Pastur distribution of random correlation matrices, the eigenvalues of correlation matrices are themselves systematically noisy when the ratio $\tau/n$ is too small. Thus, filtering the eigenvalues (without modifying the eigenvectors) also filters the correlation matrix itself. This yields so-called rotationally invariant estimators of correlation matrices (RIE).  Remarkably, how to filter only eigenvalues optimally  is known: if $n<\tau$, if the system is stationary, and if the distribution of price returns is gentle enough, the optimal RIE converges to the Oracle estimator (which knows the future realized correlation matrix)  at fixed ratio $q=\tau/n$ and in the large system limit $n$ and $\tau \to\infty$  \citep{bun2016rotational}. In practice, computing the optimal RIE is far from trivial for finite $n$, i.e., for sparse eigenvalue densities; several numerical methods address this problem, such as QuEST \citep{ledoit2012nonlinear}, Inverse Wishart regularisation \citep{bun2017cleaning}, or the cross-validated approach (CV hereafter) \citep{bartz2016cross}. 

The optimal RIE is optimal with two important restrictions: first that only the eigenvalues are filtered, and second that the system is stationary. In other words, an estimator that is more reactive, for example by allowing $\tau<n$, or that captures more of the stable structure of the true correlation matrix by also filtering the eigenvectors may be better than the optimal RIE.

A second kind of correlation (or covariance) matrix filtering rests on a well-chosen ansatz, or equivalently on a structure family of the dependence matrix. For example, linear shrinkage uses a target covariance (or correlation) matrix (it also has an eigenvalue filtering interpretation \citep{potters2005financial}). Factor models belong to the structure-based approach. A particular case is hierarchical factor models, which have been shown to yield remarkably good GMV portfolios \citep{tumminello2007hierarchically, tumminello2007kullback,pantaleo2011improved}.

A problem of the hierarchical ansatz is its sensitivity to the relatively small changes of the input data. For example, bootstrapping the original data  does not yield many statistically validated hierarchical clusters in correlation matrices of equity returns \citep{bongiorno2019nested}. Very recently,  this sensitivity was leveraged to build a more flexible estimator, which consists in averaging filtered hierarchical clustering correlation or covariance matrices over bootstraps of the input  data (BAHC) \citep{bongiorno2021covariance}.  BAHC not only allows  an imperfect hierarchical structure, i.e., a moderate overlapping among clusters, but also a probabilistic superposition of quite distinct hierarchical structures. When applied to GMV portfolios, 
BAHC  yields similar or better realized risk than the optimal eigenvalues filtering methods but for a much smaller $\tau$ than its competitors, which leads to portfolios that are much more reactive to changing market conditions. It can be further improved, as shown below.

The RIE literature, beyond  one-shot volatility estimation, can be generalized to  exponentially moving averages of correlation/covariance matrices \citep{potters2005financial,wc2021large}. The latter are a special case of dynamic conditional correlations (DCC), which is the equivalent of GARCH for correlation matrices \citep{engle2002dynamic} where the baseline correlation matrix is computed over a long time horizon and mixed with exponentially averaged point-wise estimates of correlations. This method requires $\tau>n$ points without further regularization. Convergence problems were fixed in \cite{aielli2013dynamic}.

Finally, volatility estimation and prediction are needed to obtain covariance matrices such as GARCH (or one of its variations) or rough volatility \citep{gatheral2018volatility} . \cite{engle2019largecov}  propose a way to combine both DCC, RIE, and GARCH estimation in an efficient way, which is regarded as the state of the art.

\section{Methods}

This paper extends BAHC to account for the structure of the correlation matrix that is not described by BAHC, i.e., the residuals. The rationale is that the latter may also contain  structures that persist in the out-of-sample window, hence that they should not be left out by the filtering method. The idea to filter the residuals recursively and to average the filtered matrices of bootstrapped data.

The order of recursion, denoted by $k$, is a parameter of the method, which we thus propose to call $k-$BAHC. This new method is equivalent to BAHC when $k=1$ by convention.  The higher $k$, the finer the details kept by $k-$BAHC, which, as shown below, improves the out-of-sample GMV portfolios up to a point.  When  $k$ tends to infinity, the filtered correlation matrix converges to the unfiltered correlation matrix averaged over many bootstrap copies. This matrix is almost surely positive definite in the high-dimensional regime $\tau<n$, despite the fact that the empirical unfiltered correlation matrix is not positive definite, a by-product of bootstraps, as shown by \citep{bongiorno2020bootstraps}.    

As shown below, the optimal average $k$ depends on the size of the in-sample window in a data set of US equities. It is generally an increasing function of the in-sample window length: when the latter is small, most of the variations of the empirical correlation matrices are due to estimation noise, which is best filtered by a small $k$; as  calibration window length increases, the relative importance of estimation noise decreases and thus a higher $k$ should be preferred. 

\subsection{Notations}
Let us start with some notations of standard quantities: let ${\bf R}$ be a $n \times \tau$ matrix of price returns. Its $n\times n$ covariance matrix, denoted by ${\bf \Sigma}$, has elements $\sigma_{ij}$, where
\begin{equation}
\sigma_{ij} =\frac{1}{\tau }\sum_{t=1}^\tau \left( r_{it} - \bar{r}_i \right) \left(r_{jt} - \bar{r}_j \right)
\end{equation}
and where $\bar{r}_i = \sum_{t=1}^\tau r_{it}/\tau$ is the sample mean of vector $r_i$. The Pearson correlation matrix ${\bf C}$ has elements 
\begin{equation}
c_{ij} = \frac{\sigma_{ij}}{\sqrt{\sigma_{ii}\,\sigma_{jj}}}
\end{equation}

As $k-$BAHC is an extension of BAHC, itself is a bootstrapped version of the strictly hierarchical filtering method of \cite{tumminello2007hierarchically}, let us start with hierarchical clustering.

\subsection{Hierarchical Clustering}

Hierarchical clustering agglomerates groups of objects recursively according to a distance matrix taken here as $\textbf{D}=1-\textbf{C}$ with elements $d_{ij}$; $\textbf{D}$ respects all the axioms of a proper distance. Accordingly, the distance between clusters $p$ and $q$, denoted by $\rho_{pq}$ is defined as the average distance between their elements
\begin{equation}
\rho_{pq} = \frac{\sum_{i \in \mathfrak{C}_p} \sum_{j \in \mathfrak{C}_q} d_{ij}}{n_q\,n_p},\label{eq:rho_pq}
\end{equation}
where $\mathfrak{C}_p$ and $\mathfrak{C}_q$ denote the $n_p$, respectively $n_q$, elements of clusters $p$ and $q$ respectively.

Hierarchical agglomeration works as follows: initially, each element is a cluster.. Then, the two clusters $(p,q)$ with the smallest distance $\rho_{pq}$ are merged into a new cluster $s$ which contains the elements $\mathfrak{C}_s = \mathfrak{C}_p \cup \mathfrak{C}_q$. This algorithm is applied until all the nodes form a single  cluster. This defines a tree, called a dendrogram,  which uniquely identifies the genealogy of cluster merges, denoted by $\mathfrak{G}$.

\subsection{Hierarchical Clustering Average Linkage Filtering (HCAL)}\label{sec:ALCA}
Defining a merging tree is not enough to clean correlation matrices. \cite{tumminello2007hierarchically} propose to average all the elements of the sub-correlation matrix defined from the indices $\mathfrak{F}_{pq} = \{ (i,j)\,:\, i \in \mathfrak{C}_p,\, j \in \mathfrak{C}_q\}$, i.e., to replace $c_{ij}$ by
\begin{equation}
c_{ij}^< = c_{ji}^<=1 - \rho_{pq}  \;\; \mbox{where} \;\;\;  (p,q ) \in  \mathfrak{G},\,  (i,j) \in \mathfrak{F}_{pq},
\end{equation}
where $\rho_{pq}$ is the average distance between clusters $p$ and $q$ (see Eq.~\eqref{eq:rho_pq}), with $c_{ii}^<$ set to  1. This defines the HCAL-cleaned correlation matrix ${\bf C}^<$, which corresponds to a hierarchical factor model \citep{tumminello2007hierarchically}. 

HCAL-filtered matrices have two interesting properties: by construction, ${\bf C}^<$ is positive definite when the correlation matrix is dominated by a global mode, i.e., when the average correlation is large, as in equity correlation matrices \citep{tumminello2007hierarchically}. In addition, ${\bf C}^<$ is the simplest matrix that has the same dendrogram as the empirical correlation matrix ${\bf C}$; this means that by applying the HCAL to both ${\bf C}$ and ${\bf C}^<$, the resulting dendrograms will be identical. This, however, is also one of the main limitations of this approach as it prevents any overlap among clusters; in addition, the dendrogram of ${\bf C}$ may not be the true one.

\subsection{BAHC}

\cite{bongiorno2021covariance}, noting that a single hierarchical structure may be too inflexible to describe  the dependence structure of financial correlation matrices faithfully,  introduced a bootstrapped version of HCAL, where each bootstrap may be associated to a new hierarchical structure. The recipe prescribes to create a set of $m$ bootstrap copies of the data matrix ${\bf R}$, denoted by $\{{\bf R}^{(1)},\, {\bf R}^{(2)},\, \cdots, {\bf R}^{(m)}\}$. A single bootstrap copy of the data matrix ${\bf R}^{(b)} \in \mathbb{R}^{n \times \tau}$ is defined entry-wise as $r^{(b)}_{it} = r_{i {\bf s}^{(b)}_t}$, where ${\bf{s}}^{(b)}$ is a vector of dimension $\tau$ obtained with random sampling by replacement of the elements of the vector $\{1, 2, \cdots, \tau\}$. The vectors ${\bf s}^{(b)}$, $b=1,\cdots,m$ are sampled independently.

We compute the Pearson correlation matrix ${\bf C}^{(b)}$ of each bootstrap ${\bf R}^{(b)}$ of the data matrix, from which  we derive the HCAL-filtered matrix ${\bf C}^{(b)<}$. Finally, the filtered Pearson correlation matrix ${\bf C}^{\textrm{BAHC}}$ is defined as the average over the $m$ filtered bootstrap copies, i.e.,
\begin{equation}\label{eq:corfilt}
{\bf C}^{\textrm{BAHC}} = \sum_{b=1}^m \frac{{\bf C}^{(b)<}}{m}
\end{equation}

Finally, BAHC filtered covariance is obtained from the sample univariate variance according to
\begin{equation}\label{eq:covfilt}
    \sigma_{ij}^{\textrm{BAHC}} = c_{ij}^{\textrm{BAHC}} \sigma_{ii} \sigma_{jj}
\end{equation}

Bootstraps  introduce perturbations to the original data. Whenever the hierarchical structure describes  the dependence structure of data well, bootstraps essentially lead to the same dendrogram. On the other hand, whenever a single dendrogram fails to account for the full dependence structure, bootstraps lead to several candidate dendrograms. This is why BAHC is a more flexible filtering method than HCAL.




\subsection{k-BAHC}

Let us write 
\begin{equation}
\mathbf{C}^{(b)}={\bf C}^{(b)<}+\mathbf{E}_{(1)}^{(b)}, 
\end{equation}
where $\mathbf{E}_{(1)}^{(b)}$ is the residual matrix of the HCAL filtering, which contains the dependence structure of $\mathbf{C}^{(b)}$ not included in $\mathbf{C}^{(b)<}$.   The central idea behind $k-$BAHC is to apply HCAL filtering further to $\mathbf{E}_{(1)}^{(b)}$, which yields the second order approximation of the residuals.

More generally, one write

\begin{equation}\label{eq:Ck}
{\bf C}^{(b)<}_{(k+1)} = {\bf C}^{(b)<}_{(k)} + {\bf E}^{(b)<}_{(k)}.
\end{equation}
where ${\bf E}^{(b)<}_{(k)}$ is obtained by applying HCAL filtering  to the residual matrix
\begin{equation}\label{eq:res}
{\bf E}_{(k)}^{(b)}= {\bf C}^{(b)} - {\bf C}^{(b)<}_{(k)}.
\end{equation}
with the conventions that ${\bf C}^{(b)<}_{(0)}={\bf 0}$ and ${\bf C}^{(b)<}_{(1)}={\bf C}^{(b)<}$.
When $k=0$, ${\bf E}_{(0)}^{(b)}={\bf C}^{(b)}$.
For example, $k=1$ correspond to HCAL-filtered matrix. 
The recursive application of Eqs\ \eqref{eq:Ck} and \eqref{eq:res} allows us to compute the filtered matrix at any order $k$. It is worth noticing that by iterating  Eqs\ \eqref{eq:Ck} and \eqref{eq:res},
\begin{equation}\label{eq:kinf}
\lim_{k \to \infty} {\bf C}^{(b)<}_{(k)} = {\bf C^{(b)}}
\end{equation}
as the residues become smaller and smaller. It is important to point out that ${\bf C}^<_{(k)}$ is not in general a semi-positive definite matrix for $k>1$, and in most cases, some small negative eigenvalues have been observed in our numerical experiments. These eigenvalues, according to Eq.\eqref{eq:kinf}, shrink to non-negative values when $k$ approaches infinity. For any order $k>1$, we set the possibly negative eigenvalues to 0.

As for BAHC, the filtered Pearson correlation matrix ${\bf C}^{k\textrm{-BAHC}}$ is defined as the average over the $m$ filtered bootstrap copies, i.e.,
\begin{equation}
{\bf C}^{k-\textrm{BAHC}} = \sum_{b=1}^m \frac{{\bf C}_{(k)}^{(b)<}}{m}
\end{equation}

While ${\bf C}^{(b)<}_{(k)}$ is a semi-positive definite matrix, the average of these filtered matrices rapidly becomes positive-definite, as shown in \cite{bongiorno2020bootstraps}: it is essentially a by-product of data bootstrapping. This convergence is fast, and it is guaranteed almost surely if the number of bootstraps $m \geq n$, but in most of the cases, it is reached for $m\ll n$. Finally, $k$-BAHC filtered covariance is obtained from the sample univariate variance according to
\begin{equation}
    \sigma_{ij}^{k\textrm{-BAHC}} = c_{ij}^{k\textrm{-BAHC}} \sigma_{ii} \sigma_{jj}
\end{equation}

The main advantage of $k-$BAHC over $k-$HCAL  is not to force ${\bf C}^{k-\textrm{BAHC}}$ to be embedded in a purely recursive hierarchical structure.


\section{Results}
\subsection{Data}
We consider the daily close-to-close returns from 1999-01-04 to 2020-03-31 of US equities, adjusted for dividends, splits, and other corporate events. More precisely, the data set consists of 1295 assets  taken from the union of all the components of the Russell 1000 from 2010-06 to 2020-03.  The number of stocks with data varies over time: it ranges from 497 in 1999-02-18 to 1172 in 2018-01-17.

\subsection{Spectral Properties}\label{sec:spec_prop}
Spectral properties explain why the original BAHC filtering achieves a similar or better realized variance than its competitors that focus on filtering the eigenvalues of the correlation matrix only: BAHC eigenvectors have a larger overlap with the out-of-sample eigenvectors than the unfiltered empirical eigenvectors while still filtering eigenvalues nearly as well as the optimal methods \citep{bongiorno2021covariance}.
This sub-section is devoted to investigate how the eigenvector components change as $k$ is increased. It turns out that the localization of eigenvector components is crucial in understanding the role of $k$.

To understand why localization  matters to portfolio optimization, in particular the localization of the eigenvectors associated to the smallest eigenvalues, it is worth recalling that Global Minimum Portfolios correspond to the optimal weights
\begin{equation}\label{eq:minrisk}
{\bf w}^* = \frac{{\bf \Sigma}^{-1}  \mathbb{1}}{\mathbb{1} {\bf \Sigma}^{-1}  \mathbb{1}'}
\end{equation}
which is a sum by rows (or columns) of the inverse covariance matrix,  normalized to one. The inverted covariance matrix can be expressed in terms of the spectral decomposition of ${\bf \Sigma}$ as
\begin{equation}
{\bf \Sigma}^{-1} = \sum_{i=1}^n \frac{1}{\lambda_i} {\bf v}_i {\bf v}_i',
\end{equation}
where $\lambda_i$ and ${\bf v}_i$ are respectively the $i-$th eigenvalue  of ${\bf \Sigma}$ and its associated eigenvector. This equation means that the composition of the eigenvectors related to the highest eigenvalues is irrelevant and the portfolio allocation is dominated by the eigenvectors of the smallest eigenvalues. Let us assume that the eigenvalue are ordered, i.e., that $\lambda_1>\lambda_2>\cdots>\lambda_n$. If the smallest eigenvalue is much smaller than all the others ones, i.e., $\lambda_n \ll \lambda_{n-1}$, the largest part of investment will be on the $j$-stocks such that $|v_{nj}|\gg 0$. Therefore, the localization of the non-zero elements of the  eigenvectors associated to the smallest eigenvalues is crucial to understand the portfolio allocation.

Eigenvector localization can be  summarized by the Inverse Participation Ratio (IPR), defined as
\begin{equation}
\mbox{IPR}_i =\frac{1}{ \sum_{j=1}^{n}  v_{ij}^4}
\end{equation}  
where the index $i$ refers to the $i$-th eigenvalue. The smaller the value of $\mbox{IPR}_i$, the more localized  its associated eigenvector. The most localized  case corresponding to IPR$=1$.

\begin{figure}
\centering
\subfloat[IPR cumulative distribution.\label{fig:IPRcum}]{\includegraphics[width=0.45\textwidth]{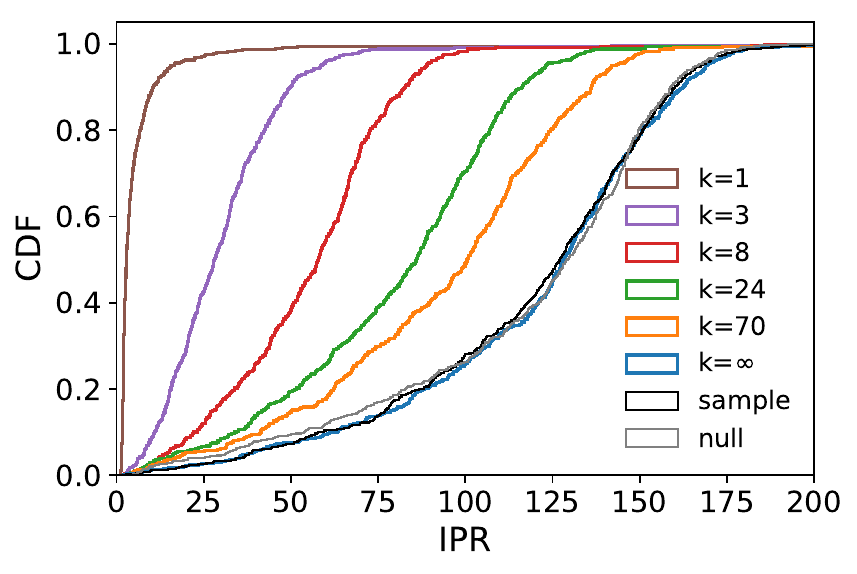} }
\subfloat[IPRs and eigenvalues scatter plot.\label{fig:IPRscat}]{\includegraphics[width=0.45\textwidth]{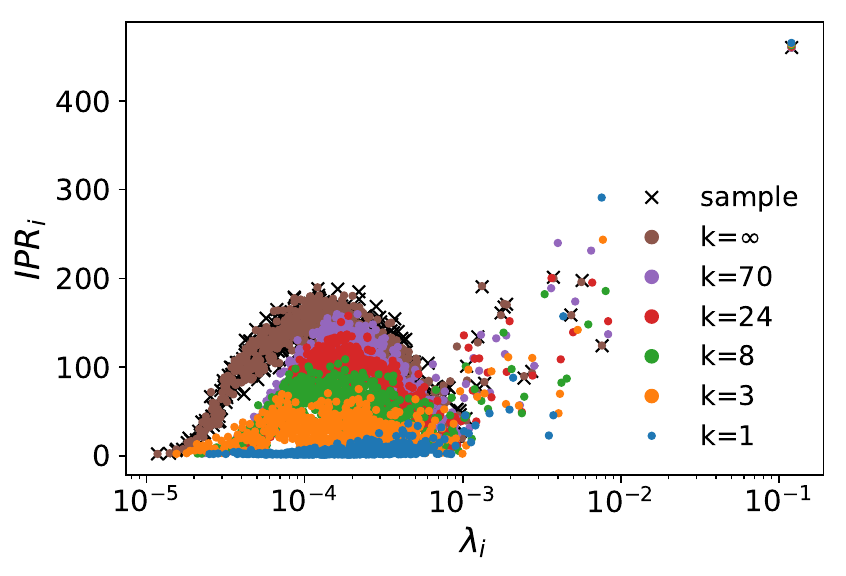}}
\caption{Left plot: cumulative distribution of the Inverse Participation Ratio (IPR) computed on $k-$BACH for different values of $k$, together with  the sample covariance IPR and the IPR of shuffled returns (null). Right plot: scatter plot of  the IPR$_i$ versus  eigenvalue $\lambda_i$. Both plots use data from the period [2016-04-12,2020-03-31] which contains 588 assets. }\label{fig:IPR}
\end{figure}

Figure \ref{fig:IPRcum} reports the cumulative distribution function of $\mbox{IPR}$ of the eigenvectors for different values of the recursion order $k$. The dependence on $k$ is obvious: $1-$BACH has the most localized eigenvectors; the larger the value of $k$, the less localized the components of the eigenvectors. In the limit  $k \to \infty$, one recovers the empirical, unfiltered, covariance matrix. In addition, the IPR of the latter two are hardly different from the random matrix null expectation obtained by shuffling price returns asset by asset in the data matrix.

Figure \ref{fig:IPRscat} gives more details about the IPRs associated with the eigenvalues. It makes it obvious that IPRs are different for small eigenvalues, while no clear pattern emerges for the outliers $\lambda_i\gtrsim  10^{-3}$. Since the lowest eigenvalues are the ones that affect mainly GVM portfolio optimization,  a filtering procedure that modifies the IPR of such eigenvalues will produce a substantial difference in the portfolio allocation.

\subsection{Global Minimum Variance portfolios}

This part explores how the realized risk of GMV portfolios depends on the recursion order $k$ and compares it with the performance obtained from sample covariance and the Cross-Validated (CV) eigenvalue shrinkage \citep{bartz2016cross}, which is a strong contender for the best realized risk \citep{bongiorno2021covariance}. Two types of tests are carried out:  because our data covers many different market regimes and a variable number of assets, we first ask what is the average realized risk of each covariance cleaning method over  random collections of assets in  randomly chosen periods of fixed length. This allows us to assess the performance of each cleaning scheme in a fair way and to control the effect of the calibration window length. In the second part, we compare the performance of these optimal portfolios with all available stocks at any given time. We also differentiate between portfolios restricted to long-only positions and those without this constraint. In both cases, there is no asset selection from the chosen universe other than that coming from the minimization of the variance.

\subsubsection{Random assets, random periods}
The experiments of this part are carried out in the following way: for each calibration window length $\tau_{in} \in [20, 2000]$ we randomly choose a time $t$ between 2000-01-03 and 2020-03-30 that defines a calibration window $[t-\tau_{in}, t[$, and a test window $[t, t+ \tau_{out}[$ with $\tau_{out}=21$ days. We then sample $n=100$ stocks over the available assets in the calibration and test windows. Finally, we compute the GMV portfolios with and without short positions using $k-$BAHC, the state-of-art Cross-Validated (CV) eigenvalue shrinkage \citep{bartz2016cross} and the unfiltered empirical covariance matrix.

\begin{figure}
\centering
\subfloat[Long-short realized risk.\label{fig:VOLlongshort}]{\includegraphics[width=0.45\textwidth]{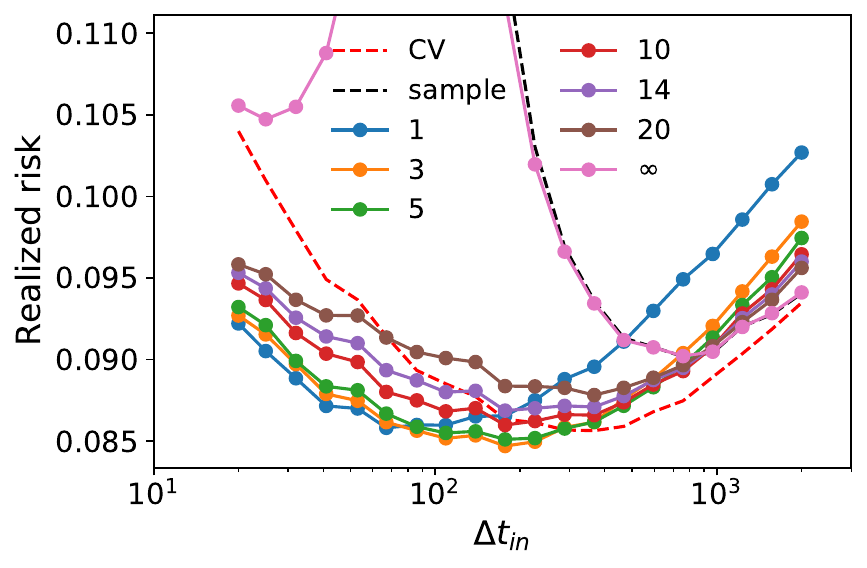}}
\subfloat[Long-short optimal $k$.\label{fig:Klongshort}]{\includegraphics[width=0.45\textwidth]{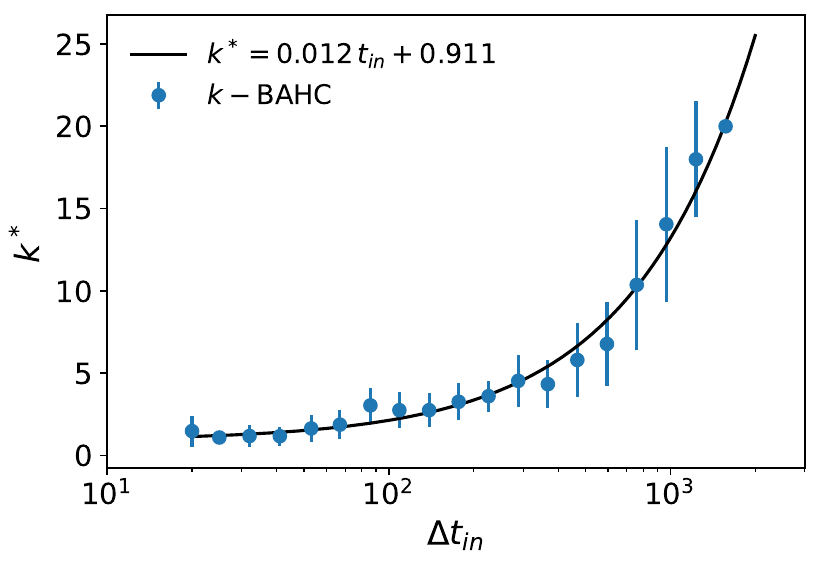}}

\subfloat[Long-only realized risk.\label{fig:VOLlong}]{
\includegraphics[width=0.45\textwidth]{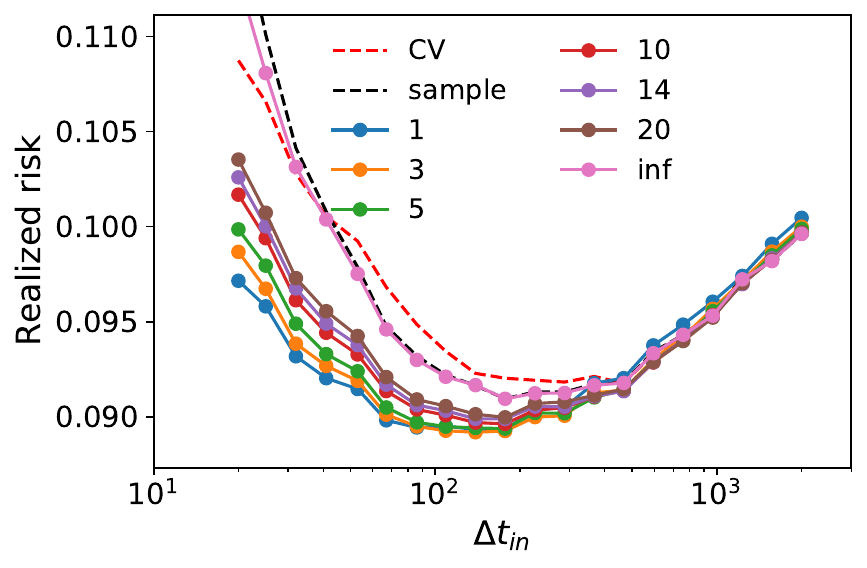}}
\subfloat[Long-only optimal $k$.\label{fig:Klong}]{
\includegraphics[width=0.45\textwidth]{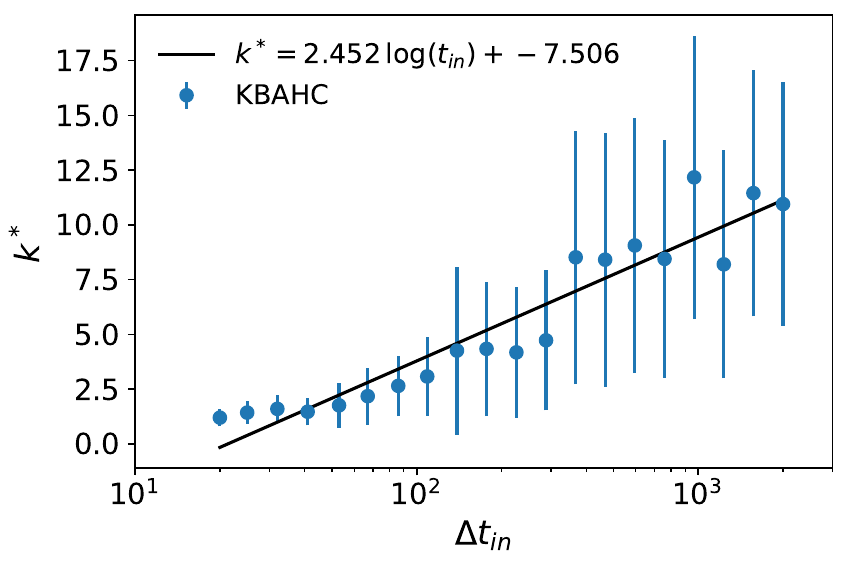}}

\caption{Left plots: annualized realized risk for different covariance estimators computed over calibration windows of length $\tau_{in}$; each point is the median of 10,000 simulations; testing period of 21 days. Legend numbers refer to  the approximation order used in $k-$BAHC. Right plots:  average optimal approximation order $k$ for different calibration windows;  the error bars  represent the standard deviation obtained by  bootstrap re-sampling of the test-period performance; the continuous black line  comes from a linear regression. The upper plots correspond to long-short portfolios and lower plots to  long-only portfolios. }\label{fig:VOL}
\end{figure}

Figure \ref{fig:VOL}  shows the realized risk of GMV portfolios obtained with the chosen filtering schemes and with the empirical (sample) covariance matrix.
The $k-$BACH estimators outperform both CV and the sample covariance estimators for $\tau_{in}<300$ in the long-short case and for every $\tau_{in}$ for the long-only case (Figures \ref{fig:VOLlongshort} and \ref{fig:VOLlong}). The highest performance of CV is obtained for $\tau_{in}\approx 350$; however, the highest absolute minimum is obtained for $k-$BAHC with $\tau_{in}\approx 200$, which requires much shorter calibration times and thus yields more reactive portfolios.

What values to take for $k$ (and for this data set) depends on $\tau_{in}$. In the high-dimensional regime ($q>1$), i.e. for  $\tau_{in}<90$, the best results are obtained for $k=1$; however, when $\tau_{in}$ increases, the performance of $k=1$ becomes even worse than the sample covariance. From this analysis, it is clear that the larger the calibration window size, the larger the approximation order $k$ must be. Figures~\ref{fig:Klongshort} and \ref{fig:Klong} show the average optimal $k^*$ that minimizes the realized risk as a function of $\tau_{in}$ for the long-short and long-only case. They confirm that a longer calibration window requires a higher approximation order both for the long-short and long-only cases; however, whereas for the long-short setting, $k^*$ seems to have a linear dependence on  $\tau_{in}$, this dependence  is sub-linear (and much noisier) in the long-only case. It is worth remarking that the fits of the right plots of Figure \ref{fig:VOL}  are obtained with $k\le20$: larger values of $k$ might further improve the performance for  larger $\tau_{in}$; however, they would require a comparatively greater computational effort. 

\subsubsection{Dynamic Conditional Covariance Estimation}
This section is devoted to a comparison between $k-$BAHC and DCC-NLS, the state-of-the-art method  to clean covariance matrices in a dynamical way \citep{engle2019largecov}: the idea behind DCC-NLS is to have a punctual estimation of the correlation matrix defined as a linear combination of a time-varying correlation matrix  
 with a  target matrix (DCC, see \citep{engle2002dynamic}) that comes from a non-linear shrinkage  (NLS) of the sample correlation matrix. The estimation of the importance of the time-varying part is obtained by likelihood maximization. Since the former approach is dynamical, although it requires a long calibration window, the final estimator is strongly influenced by the most recent history, also leading to reactive portfolios. 
The DCC is computed over devolatized returns, which are defined as the ratio between the returns and the conditional volatility, both obtained from a GARCH(1,1). The last estimation of the conditional volatility are joined into the last dynamical estimation of the correlation matrix, producing the DCC covariance matrix. 
However, since both the GARCH fitting and the likelihood estimation for the dynamical correlation matrices require typically a long time-series, e.g. $1250$ days (about 5 years), as suggested in~\cite{engle2019largecov}, the available pool of stock is necessarily limited to the ones which are simultaneously listed over the whole calibration period. This a quite important restriction since the average lifetime of an asset is about 7 years for US equities. We stress that $k$-BAHC does not have this limitation, since its reactivity comes from using very short calibration windows. In particular, as reported in the previous section, $k$-BACH performs very well also in the $\tau<n$ regime, which raises the question of the respective performance of both methods, each having distinct advantages.

\begin{figure}
\centering
\subfloat[Long-short realized risk.\label{fig:DCCrisklongshort}]{\includegraphics[width=0.45\textwidth]{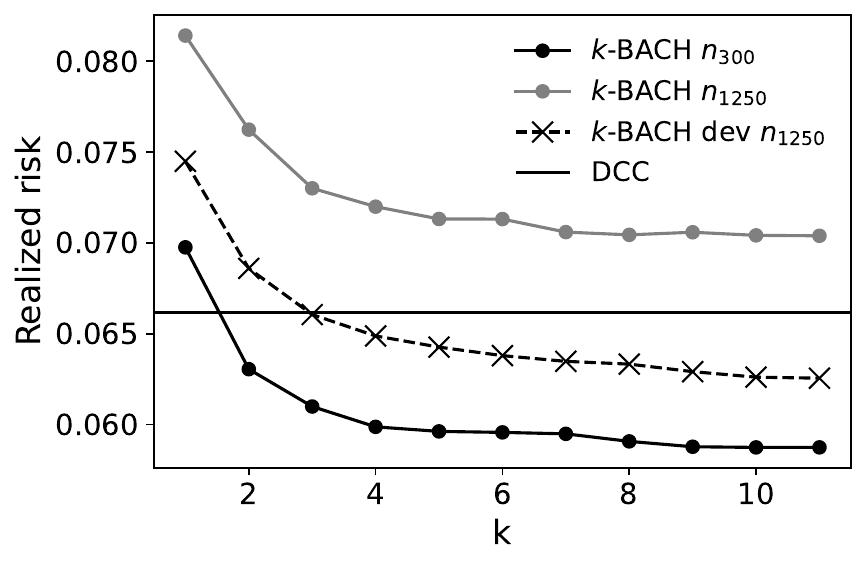}}
\subfloat[Long-short fraction of times $k$-BAHC performs.\label{fig:DCCdeltalongshort}]{\includegraphics[width=0.45\textwidth]{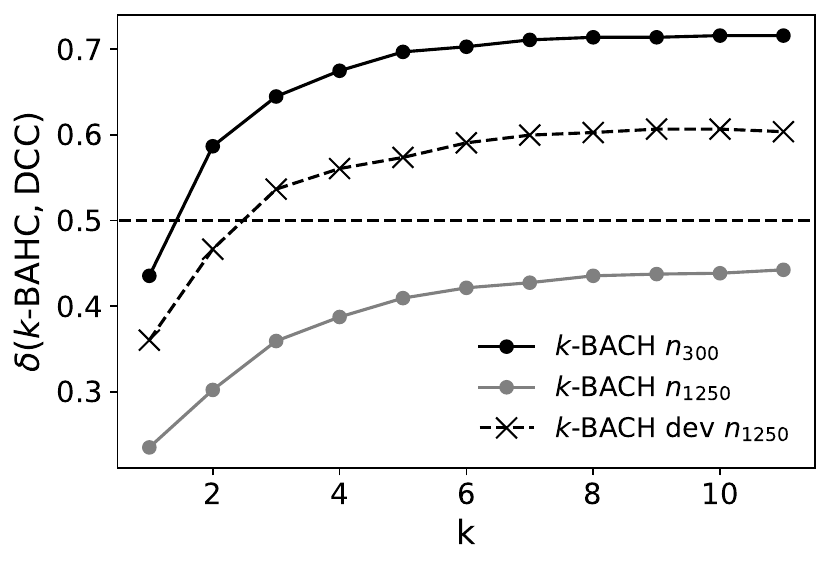}}

\subfloat[Long-only realized risk.\label{fig:DCCrisklong}]{\includegraphics[width=0.45\textwidth]{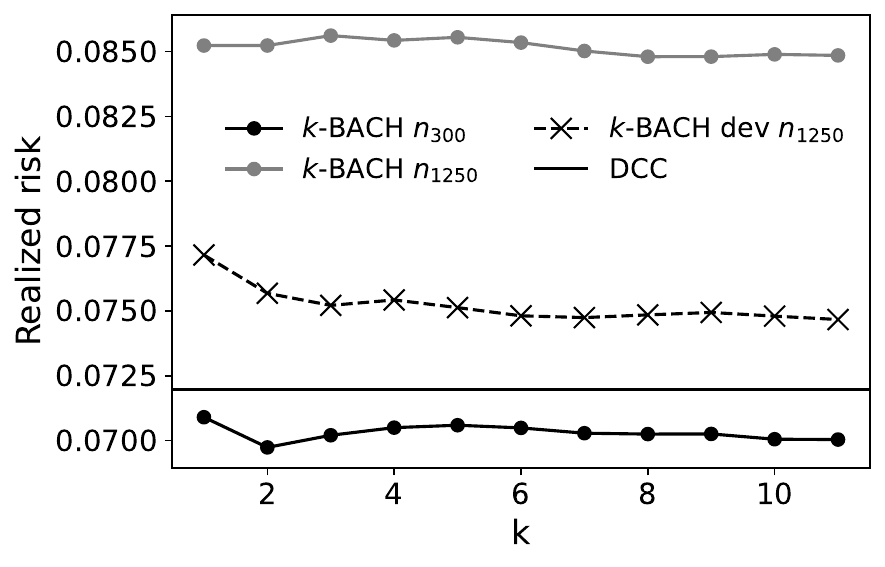}}
\subfloat[Long-only fraction of times $k$-BAHC performs.\label{fig:DCCdeltalong}]{\includegraphics[width=0.45\textwidth]{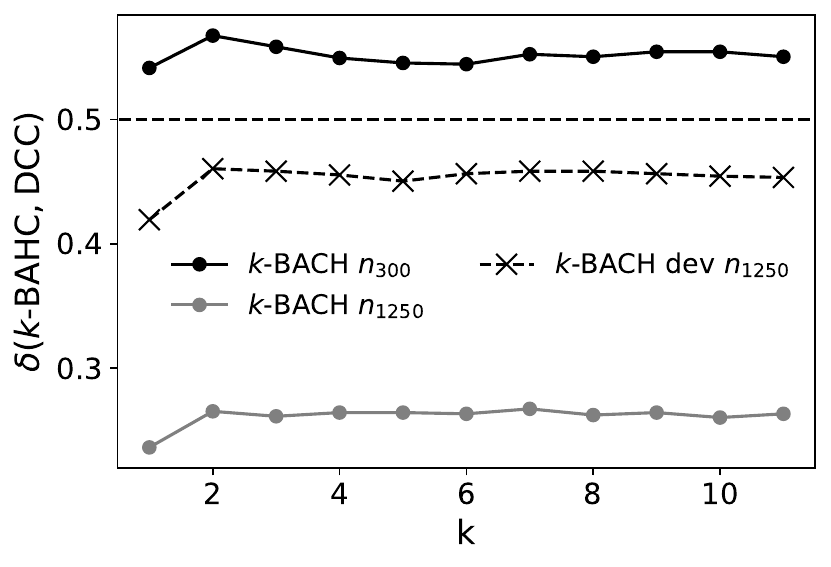}}
\caption{Left plots:  median realized risk over the next 21 days of out-of-sample of $k$-BAHC and DCC (solid line). Right plots:  fraction of times $k$-BAHC reaches a lower realized risk than DCC. Upper plots refer to long-short portfolios, lower plots refers to long-only portfolios. All the quantities are plotted against the filtering recursion level $k$ of $k-$BAHC.  For $k$-BAHC we considered three cases: 1) $n_{300}$: the portfolio is obtained by considering all available stocks in a time-window of $300$ days; 2) $n_{1250}$: the portfolio is obtained by restricting the available stocks to the ones considered by DCC, which are all the stocks available in over the whole time-window of the previous $1250$ days;  3) $n_{1250}$ dev: the portfolio is obtained by restricting the available stocks to the ones considered by DCC, and the returns are devolatized. In all the cases, the $k$-BAHC covariance estimator is computed using data from last $300$ days. All simulations consist in $1,000$ independent simulations obtained by selecting the first out-of-sample day within [2000-01-02,2020-03-31]. The lower number of simulations in this figure is forced by the substantial computation time required by DCC.}\label{fig:DCC}
\end{figure}

In order to perform this comparison, we used the {\tt R} implementation of DCC from~\cite{package:DCC}. For each simulation we randomly selected  the first day of the out-of-sample period in the [2000-01-02,2020-03-31] period and we selected all the stocks were listed in the previous 1250 days for DCC and all the stocks that were listed in the previous 300 days for $k$-BACH, denoted respectively by $n_{1250}$ and $n_{300}$. The larger pool of stocks for $n_{300}$ gives more diversification opportunities, which leads in general to lower realized risk of GMV portfolios. In Fig.~\ref{fig:DCC}, we show that $k$-BAHC outperforms DCC when $k>1$, reaching its optimal performance in the long-short case for $k>5$. In the long-only case, the gain of $k$-BAHC is approximately constant for all the  values of $k$ explored here. In particular, in Fig.~\ref{fig:DCCdeltalongshort}  and Fig.~\ref{fig:DCCdeltalong}, we show the fraction of times $k$-BACH obtains a lower risk than DCC, which reaches 72\% and 57\% for the long-short and long-only respectively (Fig 3b). 

In order to have a fair comparison and to show that this gain mainly comes  from the larger pool of available assets, we include also a set of simulations where the available pool of assets for $k$-BAHC is restricted to the ones used by DCC, i.e., $n_{1250}$, in this case, DCC outperforms $k$-BAHC  for small values of $k$ in the long-short case, although $k$-BAHC reaches similar performances for large $k$. On the other hand, the long-only portfolios are totally dominated by DCC. 
To understand better the respective role of GARCH and DCC,  we also apply $k-$BAHC to the devolatized returns of the set of assets used by DCC, i.e., $n_{1250}$. This time, $k-$BAHC outperforms DCC in the long-short case for $k\ge3$ (Fig. 3c). For the long-only case, even if the performance of $k-$BAHC significantly increases, DCC is still better (Fig. 3d). 

\subsubsection{Full-universe, full period backtest}
In this section we performed a set of portfolio optimizations with monthly computations of new portfolio weights (and rebalancing) over the full time-period [2000-01-02, 2020-03-31] for all the considered covariance estimators and  equally weighted portfolios (EW hence after), and for different in-sample window lengths. The backtests include transaction costs set to 2 bps. A slight complication comes from the variable number of available assets. Thus, as any time, $q=n/\tau$  varies and generally increases with time at fixed calibration window length. In any case, it is worth keeping in mind that $n$ is relatively large, i.e., between 497 and 1172. 

 In particular, at each rebalancing time-step we considered all the available stocks listed in both the in-sample and out-of-sample periods. 
The present work investigates in detail short calibration windows: we chose a sequence of $\tau_{in} \in [21,252]$ days by steps of $21$ days, i.e., about  $1,2, \cdots, 12$ months. We emphasize that this corresponds to $n>\tau$. In addition, for sake of completeness we also included longer calibration windows $\tau_{in} = 300, 350, 400, 500, 750, 1000, 1500, 2000$.
For each method and calibration window, we computed the realized annualized volatility, the realized annualized return, the Sharpe ratio, the gross-leverage, the concentration of the portfolio, and the average turnover. 

\begin{figure}
\centering
\subfloat[Long-short]{\includegraphics[width=0.95\textwidth]{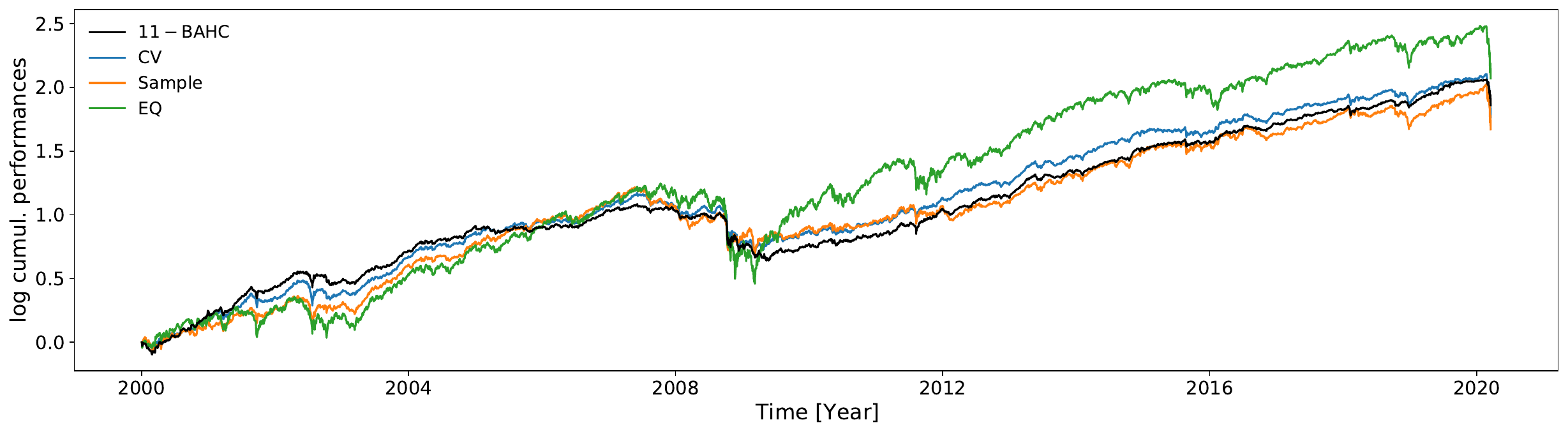}  }

\subfloat[Long-only]{\includegraphics[width=0.95\textwidth]{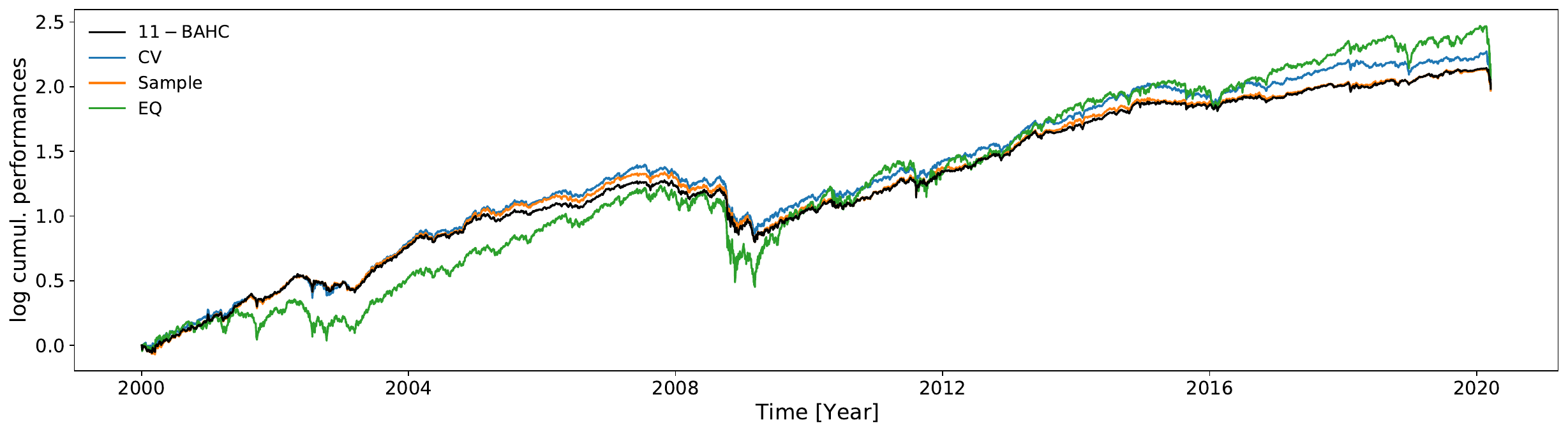} }
\caption{Cumulative performances obtained with 21 days between weights updates and rebalancing for different methods.  Both plots were obtained with a calibration window of 105 days (5 months). The upper plot refers to long-short portfolios, the lower one to long-only portfolios}\label{fig:profit}
\end{figure}

Figure \ref{fig:profit} shows the behavior of equally weighted portfolios and GMV portfolios obtained by $11-$BAHC, the globally best value for the data set that we used. As expected, GMV reduces the realized risk with respect to EW portfolios, and cleaning the covariance matrix is clearly beneficial as well.

Let us start with realized risk, the focus of this paper. The realized risk of EQ portfolios is much larger than that resulting from the other methods, which is hardly surprising as the latter account for the covariance matrix (Tables \ref{tab:SDlongshort} and \ref{tab:SDlong});  $k-$BACH achieves the smallest realized risk, and the best value for $k$ increases as $\tau_{in}$ increases.


Although GMV does not guarantee a positive return, we also report the Sharpe ratios of the various filtering methods. Because computing Sharpe ratios with moments is not efficient for heavy-tailed variables, we use the efficient and unbiased moment-free SR estimator introduced in \cite{challet2017sharper} and implemented in \cite{package:sharperRatio}. Sharpe ratios paint a picture similar to realized risk (see Tables \ref{tab:SRlongshort} and \ref{tab:SRlong}):  $k-$BAHC outperforms all the other methods for medium to large values of $k$ for almost every $\tau_{in}$ both in the long-only and long-short cases, especially when the calibration window is smaller than a year, which corresponds to $q=n/\tau\in[2,4.5]$, i.e., quite deep in the high-dimensional regime. In particular, in the long-short case, the highest SR equals $1.25$ for $k= 7, 11, 18$ in the remarkably short calibration window $\tau_{in}=105$ days (about 5 months),  and is significantly higher than the best performance of CV (SR$=1.18$) obtained for a much higher calibration window $\tau_{in}=400$. This shows that reactive portfolio optimization is invaluable. In the long-only case, the improvement of $k-$BAHC is smaller: the best SR (1.13) is that of  $18-$BAHC, whereas the highest SR of CV is $1.07$. 

However, as shown from the cumulative performances in Figure \ref{fig:profit}, the relative performance changes over time. To overcome this limitation, we evaluated the SR every year, and we performed a dense ranking of all the methods after having rounded the related SRs to the second decimal (to ensure some equal ranks). Finally we associated a score, denoted by $\langle \mbox{rank} \rangle$, to every method defined as the average dense rank over the years.  The results for the long-short and long-only cases are summarized in Table \ref{tab:rank}. It is worth noting that medium to large values of $k$ for $k-$BACH outperform all the other methods and that the optimal performance is achieved with a calibration window length shorter than for CV by a factor of about four. 

We checked that the portfolios obtained with $k-$BAHC are more concentrated than the other ones, which is consistent with the fact that the IPR of the relevant eigenvectors is smaller. The concentration of a portfolio can be measured with
\begin{equation}
n_{\textrm{eff}} = \frac{1}{\sum_{i=1}^n w_i^2}
\end{equation}
as proposed in \cite{bouchaud2003theory}; However, as noticed in \cite{pantaleo2011improved}, this quantity does not have a clear interpretation when short selling is allowed. To overcome this issue, \cite{pantaleo2011improved} introduced the $n_{90}$ metrics which measures the smallest number of stocks that amount for at least 90\% of the invested capital. Accordingly, we used $n_{90}$ is for the long-short case and $n_{\textrm{eff}}$ for the long-only one.  Looking at Tables \ref{tab:N90longshort} and \ref{tab:Nefflong}, the number of stocks selected is systematically smaller for every $k$ and calibration window for $k-$BAHC for both long-only and long-short portfolios.

That said,  $k-$BAHC has two drawbacks. First, the gross leverage is generally larger than for CV in the long-short case (see Table~\ref{tab:Grosslongshort}).  However, if we compare the values of gross leverage corresponding to the larger SR for CV and $k-$BAHC for $\tau_{in}$ within one year, they differ only by $0.52$ ($2.47$ for CV and $3.00$ for $k-$BAHC). On the other hand, without constraining the calibration window, the highest SR for CV is achieved for $\tau_{in}=400$ and the gross leverage reaches $3.31$, which is larger than for other methods. 

The other drawback of $k-$BAHC is that it requires a larger turnover for  long-short portfolios. A natural turnover metrics, denoted by  $\gamma$, was defined in \cite{reigneron2020agnostic} as
\begin{equation}
\gamma = \frac{1}{\theta } \sum_{h=0}^{\theta-1} \sum_{i=1}^n \left| w_i(t_0+h \tau_{out} ) -w_i(t_0+(h+1)\tau_{out}) \right|,
\end{equation}
where $\theta$ is the number of rebalancing operations and $t_0$ is the initial time; $\gamma$ measures the average changes in the portfolio allocation between two consecutive portfolio allocations. Table \ref{tab:Gammalongshort} shows that  $k-$BAHC has a  $\gamma$  typically twice as large as CV, except for $k=1$ and large $\tau_{in}$ for  long-short portfolios. For the long-only case (Table \ref{tab:Gammalong}) CV still outperforms $k-$BAHC in that respect, although not by much. All performance measures take into account into account the rebalancing costs. Note that the larger turnover comes from the fact that  portfolios are more concentrated, i.e., select fewer assets. It is therefore more likely that the set of selected assets changes at every weight updates.

\section{Discussion}

By combining recursive hierarchical clustering average linkage and bootstrapping of the data matrix yields a globally better way to filtering asset price covariance matrices. We have shown that this method filters the eigenvectors associated with small eigenvalues of the covariance matrix by making them more concentrated, which in turn yields portfolios with fewer assets. Because $k-$ BACH captures more of the persistent structure of covariance matrices with shorter calibration windows, it produces Global Minimum Variance portfolios with smaller realized volatility  than even the optimal Rotationally Invariant Estimators. Furthermore, because it works well for very short calibration windows, it is well suited to evolving investment universes and changing market conditions, and thus can profit from more diversification opportunities than conditional methods that require much longer timeseries. This is why an unconditional covariance approach with $k-$BAHC filtering outperforms state-of-the-art conditional volatility methods for long-short portfolios.

The main drawback of $k-$BAHC is that it requires a larger turnover. This is due, in part, to the fact that resulting portfolios are more concentrated, hence that the fraction of capital in which to invest change more rapidly than less specific methods. Whether this reflects a genuine change of market structure or a by-product of the specific assumptions of $k-$BACH is an interesting open question.

Future work will investigate how $k-$BAHC may improve other kinds of portfolio optimization schemes and other financial applications of covariance matrices. A detailed exploration of the performances of $k$-BAHC on devolatized returns with the DCC approach will be also addressed in future works.

\section*{Acknowledgement(s)}

 This work was performed using HPC resources from the ``M\'esocentre'' computing center of CentraleSup\'elec and \'Ecole Normale Sup\'erieure Paris-Saclay supported by CNRS and R\'egion \^{I}le-de-France (\url{http://mesocentre.centralesupelec.fr/})

\section*{Funding}

This publication stems from a partnership between CentraleSup\'elec and BNP Paribas.



\bibliography{report_kbahc}

\newpage

\begin{table}
\centering
\begin{tabular}{lccc}
\toprule
& \multicolumn{2}{l}{Long-short } \\ \cmidrule{2-4}
rank &  $\langle$ rank $\rangle$ &     Method &  $\tau_{in}$ \\
\midrule
1    &                     39.00 &  $11-$BAHC &              105 \\
2    &                     39.48 &  $30-$BAHC &               84 \\
3    &                     39.76 &  $18-$BAHC &               63 \\
4    &                     40.10 &  $18-$BAHC &              105 \\
5    &                     40.14 &  $11-$BAHC &               84 \\
6    &                     40.57 &  $30-$BAHC &               63 \\
7    &                     41.00 &  $18-$BAHC &               84 \\
8    &                     41.38 &  $11-$BAHC &               63 \\
9    &                     42.14 &   $7-$BAHC &               84 \\
10   &                     42.57 &  $30-$BAHC &              105 \\
$\cdots$  &           $\cdots$ &  $\cdots$ &              $\cdots$ \\
31   &                     47.62 &         CV &              400 \\
143  &                     65.24 &     Sample &             1500 \\
192  &                     81.57 &         EQ &              - \\
\bottomrule
\end{tabular}
\hspace{40.pt}
\begin{tabular}{lccc}
\toprule
& \multicolumn{2}{l}{Long-only } \\ \cmidrule{2-4}
rank &  $\langle$ rank $\rangle$ &     Method &  $\tau_{in}$ \\
\midrule
1   &                     36.57 &  $11-$BAHC &               63 \\
2   &                     38.43 &  $30-$BAHC &               63 \\
3   &                     39.62 &   $7-$BAHC &               63 \\
4   &                     40.90 &  $18-$BAHC &               63 \\
5   &                     42.29 &  $30-$BAHC &               84 \\
6   &                     42.67 &  $11-$BAHC &              126 \\
7   &                     42.86 &   $4-$BAHC &               63 \\
8   &                     43.57 &   $4-$BAHC &              126 \\
9   &                     43.95 &   $3-$BAHC &               63 \\
10  &                     44.19 &  $11-$BAHC &              105 \\
$\cdots$  &           $\cdots$ &  $\cdots$ &              $\cdots$ \\
20  &                     46.76 &         CV &              126 \\
29  &                     48.52 &     Sample &              400 \\
168 &                     73.86 &         EQ &               - \\
\bottomrule
\end{tabular}

\vspace{10.pt}
\caption{Average rank of Sharpe ratios computed year by year, denoted by $\langle \mbox{rank}\rangle$, of the various methods for different in-sample window sizes $\tau_{in}$. The left table refers to long-short portfolios, and the right table to long-only portfolios.}
\label{tab:rank}
\end{table}

\setcounter{table}{1}
\begin{sidewaystable}

\begin{tabular}{lccccccccccc}
\toprule
& \multicolumn{2}{l}{Covariance matrix estimators } \\ \cmidrule{2-12}
$\tau_{in}$&    EQ &  Sample &    CV &  $1-$BAHC &  $2-$BAHC &  $3-$BAHC &  $4-$BAHC &  $7-$BAHC &  $11-$BAHC &  $18-$BAHC &  $30-$BAHC \\\midrule
21   & 0.202 &     NaN & 0.123 &     0.101 &     0.101 &\textbf{0.100}&\textbf{0.100}&     0.101 &      0.102 &      0.102 &      0.103  \\
42   & 0.203 &     NaN & 0.113 &     0.098 &     0.097 &\textbf{0.096}&\textbf{0.096}&\textbf{0.096}&\textbf{0.096}&\textbf{0.096}&      0.097  \\
63   & 0.204 &   0.146 & 0.107 &     0.091 &     0.088 &\textbf{0.087}&\textbf{0.087}&\textbf{0.087}&\textbf{0.087}&      0.088 &      0.088  \\
84   & 0.204 &   0.134 & 0.103 &     0.091 &     0.086 &     0.085 &     0.085 &\textbf{0.084}&\textbf{0.084}&      0.085 &      0.085  \\
105  & 0.204 &   0.129 & 0.100 &     0.091 &     0.086 &     0.084 &     0.084 &\textbf{0.083}&      0.084 &      0.084 &      0.084  \\
126  & 0.204 &   0.124 & 0.099 &     0.092 &     0.086 &     0.084 &\textbf{0.083}&\textbf{0.083}&\textbf{0.083}&\textbf{0.083}&\textbf{0.083} \\
147  & 0.204 &   0.120 & 0.098 &     0.092 &     0.085 &     0.083 &     0.083 &\textbf{0.082}&\textbf{0.082}&\textbf{0.082}&      0.083  \\
168  & 0.204 &   0.118 & 0.097 &     0.093 &     0.086 &     0.083 &     0.083 &\textbf{0.082}&\textbf{0.082}&\textbf{0.082}&\textbf{0.082} \\
189  & 0.204 &   0.116 & 0.096 &     0.094 &     0.086 &     0.083 &     0.082 &     0.082 &\textbf{0.081}&      0.082 &      0.082  \\
210  & 0.203 &   0.116 & 0.095 &     0.095 &     0.087 &     0.084 &     0.083 &\textbf{0.082}&\textbf{0.082}&\textbf{0.082}&\textbf{0.082} \\
231  & 0.203 &   0.114 & 0.095 &     0.100 &     0.091 &     0.088 &\textbf{0.087}&\textbf{0.087}&\textbf{0.087}&\textbf{0.087}&\textbf{0.087} \\
252  & 0.203 &   0.113 & 0.094 &     0.101 &     0.092 &     0.089 &     0.088 &\textbf{0.087}&\textbf{0.087}&\textbf{0.087}&\textbf{0.087} \\
\midrule 
300  & 0.203 &     NaN & 0.092 &     0.101 &     0.092 &     0.088 &     0.087 &\textbf{0.086}&\textbf{0.086}&\textbf{0.086}&\textbf{0.086} \\
350  & 0.203 &     NaN & 0.091 &     0.103 &     0.093 &     0.089 &     0.088 &\textbf{0.086}&\textbf{0.086}&\textbf{0.086}&\textbf{0.086} \\
400  & 0.203 &     NaN & 0.092 &     0.105 &     0.094 &     0.091 &     0.089 &     0.088 &\textbf{0.087}&\textbf{0.087}&      0.088  \\
500  & 0.202 &     NaN & 0.091 &     0.107 &     0.096 &     0.092 &     0.090 &\textbf{0.088}&\textbf{0.088}&\textbf{0.088}&\textbf{0.088} \\
750  & 0.201 &     NaN & 0.092 &     0.110 &     0.099 &     0.095 &     0.093 &\textbf{0.091}&\textbf{0.091}&\textbf{0.091}&\textbf{0.091} \\
1000 & 0.200 &   0.207 &\textbf{0.093}&     0.113 &     0.103 &     0.099 &     0.097 &     0.095 &      0.094 &\textbf{0.093}&\textbf{0.093} \\
1500 & 0.197 &   0.102 &\textbf{0.096}&     0.119 &     0.109 &     0.104 &     0.102 &     0.099 &      0.098 &      0.098 &      0.098  \\
2000 & 0.196 &   0.105 &\textbf{0.102}&     0.125 &     0.115 &     0.110 &     0.109 &     0.107 &      0.106 &      0.105 &      0.104  \\\bottomrule
\end{tabular}

\caption{Realized annualized risk of the long-short portfolio; rebalancing every 21 days. Bold entries are the optimal values for each $\tau_{in}$.}\label{tab:SDlongshort}
\end{sidewaystable}

\setcounter{table}{2}
\begin{sidewaystable}
\begin{tabular}{lccccccccccc}
\toprule
& \multicolumn{2}{l}{Covariance matrix estimators } \\ \cmidrule{2-12}
$\tau_{in}$&    EQ &  Sample &    CV &  $1-$BAHC &  $2-$BAHC &  $3-$BAHC &  $4-$BAHC &  $7-$BAHC &  $11-$BAHC &  $18-$BAHC &  $30-$BAHC \\\midrule
21   & 0.202 &   0.151 & 0.133 &\textbf{0.110}&\textbf{0.110}&     0.111 &     0.111 &     0.112 &      0.113 &      0.114 &      0.115  \\
42   & 0.203 &   0.127 & 0.122 &     0.108 &\textbf{0.107}&     0.108 &     0.108 &     0.108 &      0.109 &      0.110 &      0.111  \\
63   & 0.204 &   0.113 & 0.117 &     0.101 &\textbf{0.100}&\textbf{0.100}&     0.101 &     0.101 &      0.102 &      0.103 &      0.103  \\
84   & 0.204 &   0.107 & 0.115 &     0.100 &\textbf{0.099}&\textbf{0.099}&     0.100 &     0.100 &      0.101 &      0.101 &      0.102  \\
105  & 0.204 &   0.107 & 0.113 &     0.101 &\textbf{0.100}&\textbf{0.100}&\textbf{0.100}&     0.101 &      0.101 &      0.102 &      0.102  \\
126  & 0.204 &   0.103 & 0.110 &     0.099 &\textbf{0.098}&\textbf{0.098}&\textbf{0.098}&     0.099 &      0.099 &      0.099 &      0.099  \\
147  & 0.204 &   0.101 & 0.109 &     0.097 &\textbf{0.096}&\textbf{0.096}&\textbf{0.096}&     0.097 &      0.097 &      0.097 &      0.098  \\
168  & 0.204 &   0.100 & 0.109 &     0.097 &\textbf{0.096}&     0.097 &     0.097 &     0.097 &      0.097 &      0.098 &      0.098  \\
189  & 0.204 &   0.100 & 0.109 &     0.097 &\textbf{0.096}&     0.097 &     0.097 &     0.097 &      0.098 &      0.098 &      0.098  \\
210  & 0.203 &   0.101 & 0.109 &\textbf{0.098}&\textbf{0.098}&\textbf{0.098}&\textbf{0.098}&\textbf{0.098}&      0.099 &      0.099 &      0.099  \\
231  & 0.203 &   0.105 & 0.109 &     0.104 &\textbf{0.103}&\textbf{0.103}&\textbf{0.103}&\textbf{0.103}&\textbf{0.103}&      0.104 &      0.104  \\
252  & 0.203 &   0.106 & 0.109 &     0.105 &\textbf{0.103}&\textbf{0.103}&\textbf{0.103}&     0.104 &      0.104 &      0.104 &      0.104  \\
\midrule 
300  & 0.203 &   0.105 & 0.108 &     0.105 &\textbf{0.103}&\textbf{0.103}&\textbf{0.103}&     0.104 &      0.104 &      0.104 &      0.104  \\
350  & 0.203 &   0.106 & 0.109 &     0.105 &\textbf{0.104}&\textbf{0.104}&\textbf{0.104}&\textbf{0.104}&\textbf{0.104}&      0.105 &      0.105  \\
400  & 0.203 &   0.107 & 0.110 &     0.107 &\textbf{0.105}&\textbf{0.105}&\textbf{0.105}&     0.106 &      0.106 &      0.106 &      0.106  \\
500  & 0.202 &   0.109 & 0.111 &     0.109 &\textbf{0.107}&\textbf{0.107}&\textbf{0.107}&\textbf{0.107}&\textbf{0.107}&      0.108 &      0.108  \\
750  & 0.201 &   0.114 & 0.115 &     0.114 &\textbf{0.113}&\textbf{0.113}&\textbf{0.113}&\textbf{0.113}&\textbf{0.113}&\textbf{0.113}&\textbf{0.113} \\
1000 & 0.200 &   0.118 & 0.119 &     0.119 &     0.118 &     0.118 &     0.118 &     0.118 &      0.118 &      0.118 &\textbf{0.117} \\
1500 & 0.197 &   0.121 & 0.123 &     0.123 &     0.122 &     0.121 &     0.121 &     0.121 &      0.121 &      0.121 &\textbf{0.120} \\
2000 & 0.196 &\textbf{0.124}& 0.126 &     0.128 &     0.125 &\textbf{0.124}&\textbf{0.124}&\textbf{0.124}&\textbf{0.124}&\textbf{0.124}&\textbf{0.124} \\\bottomrule
\end{tabular}

\caption{Realized annualized risk of the long-only portfolios; rebalancing every 21 days. Bold entries are the optimal values for each $\tau_{in}$.}
\label{tab:SDlong}
\end{sidewaystable}

\setcounter{table}{3}
\begin{sidewaystable}
\begin{tabular}{lccccccccccc}
\toprule
& \multicolumn{2}{l}{Covariance matrix estimators } \\ \cmidrule{2-12}
$\tau_{in}$&   EQ &  Sample &   CV &  $1-$BAHC &  $2-$BAHC &  $3-$BAHC &  $4-$BAHC &  $7-$BAHC &  $11-$BAHC &  $18-$BAHC &  $30-$BAHC \\\midrule
21   & 0.57 &     NaN & 0.76 &      0.76 &      0.73 &      0.78 &      0.81 &      0.82 &       0.83 &       0.83 &\textbf{0.84} \\
42   & 0.61 &     NaN & 0.80 &      0.81 &      0.81 &      0.86 &      0.86 &      0.92 &       0.93 &\textbf{0.94}&       0.91  \\
63   & 0.55 &    0.75 & 0.95 &      0.91 &      1.02 &      1.09 &      1.13 &      1.19 &\textbf{1.20}&       1.19 &       1.17  \\
84   & 0.55 &    0.75 & 0.98 &      0.94 &      1.10 &      1.15 &      1.18 &      1.22 &\textbf{1.24}&       1.22 &       1.23  \\
105  & 0.54 &    0.75 & 1.06 &      0.99 &      1.10 &      1.17 &      1.19 &\textbf{1.25}&\textbf{1.25}&\textbf{1.25}&       1.23  \\
126  & 0.54 &    0.81 & 1.06 &      0.90 &      1.00 &      1.07 &      1.08 &\textbf{1.16}&       1.15 &\textbf{1.16}&\textbf{1.16} \\
147  & 0.55 &    0.84 & 1.07 &      0.89 &      0.99 &      1.04 &      1.10 &      1.19 &\textbf{1.20}&       1.19 &       1.19  \\
168  & 0.53 &    0.84 & 1.02 &      0.86 &      0.96 &      1.02 &      1.04 &      1.11 &\textbf{1.12}&       1.11 &       1.10  \\
189  & 0.53 &    0.78 & 1.06 &      0.85 &      1.01 &      1.01 &      1.07 &      1.10 &       1.16 &       1.16 &\textbf{1.17} \\
210  & 0.55 &    0.77 & 1.02 &      0.88 &      1.01 &      1.06 &      1.09 &      1.15 &       1.14 &\textbf{1.16}&       1.15  \\
231  & 0.52 &    0.71 & 1.08 &      0.82 &      0.97 &      1.03 &      1.05 &      1.09 &\textbf{1.10}&\textbf{1.10}&       1.09  \\
252  & 0.52 &    0.57 & 1.07 &      0.82 &      0.93 &      0.99 &      0.97 &      1.04 &       1.07 &\textbf{1.09}&       1.07  \\
\midrule 
300  & 0.52 &     NaN &\textbf{1.16}&      0.83 &      0.99 &      1.02 &      1.05 &      1.08 &       1.08 &       1.12 &       1.10  \\
350  & 0.51 &     NaN & 1.15 &      0.86 &      0.97 &      1.01 &      1.05 &      1.11 &       1.12 &       1.11 &\textbf{1.16} \\
400  & 0.52 &     NaN &\textbf{1.18}&      0.87 &      0.96 &      1.05 &      1.09 &      1.12 &       1.13 &       1.15 &       1.13  \\
500  & 0.50 &     NaN &\textbf{1.14}&      0.84 &      1.01 &      1.06 &      1.09 &      1.13 &\textbf{1.14}&       1.13 &       1.12  \\
750  & 0.51 &     NaN &\textbf{1.08}&      0.86 &      1.01 &      1.03 &\textbf{1.08}&\textbf{1.08}&       1.07 &       1.07 &       1.04  \\
1000 & 0.50 &    0.10 & 1.06 &      0.87 &      0.96 &      1.05 &      1.09 &\textbf{1.14}&       1.13 &       1.12 &       1.08  \\
1500 & 0.46 &    1.00 & 1.21 &      0.90 &      1.07 &      1.13 &      1.17 &\textbf{1.22}&       1.21 &       1.21 &       1.21  \\
2000 & 0.45 &    1.06 &\textbf{1.10}&      0.86 &      0.98 &      0.97 &      0.99 &      1.04 &       1.06 &       1.09 &\textbf{1.10} \\\bottomrule
\end{tabular}

\caption{Sharpe ratio of the long-short portfolios; rebalancing every 21 days. Bold entries are the optimal values for each $\tau_{in}$.}\label{tab:SRlongshort}
\end{sidewaystable}

\setcounter{table}{4}
\begin{sidewaystable}
\begin{tabular}{lccccccccccc}
\toprule
& \multicolumn{2}{l}{Covariance matrix estimators } \\ \cmidrule{2-12}
$\tau_{in}$&   EQ &  Sample &   CV &  $1-$BAHC &  $2-$BAHC &  $3-$BAHC &  $4-$BAHC &  $7-$BAHC &  $11-$BAHC &  $18-$BAHC &  $30-$BAHC \\\midrule
21   & 0.60 &    0.53 & 0.74 &      0.73 &      0.76 &      0.77 &      0.79 &      0.80 &       0.78 &       0.78 &\textbf{0.81} \\
42   & 0.58 &    0.84 & 0.81 &      0.77 &      0.84 &      0.91 &      0.91 &      0.93 &       0.95 &       0.96 &\textbf{0.97} \\
63   & 0.54 &    1.00 & 0.89 &      0.97 &      1.00 &      1.05 &      1.06 &      1.09 &\textbf{1.12}&       1.11 &       1.09  \\
84   & 0.57 &\textbf{1.05}& 0.92 &      0.94 &      0.95 &      0.98 &      1.00 &      1.02 &       1.02 &       1.02 &       1.04  \\
105  & 0.54 &    0.95 & 0.93 &      0.90 &      0.93 &      0.97 &      1.01 &      1.03 &       1.04 &\textbf{1.07}&       1.06  \\
126  & 0.57 &    1.05 & 1.01 &      0.98 &      1.01 &      1.08 &      1.09 &      1.10 &       1.11 &\textbf{1.13}&       1.12  \\
147  & 0.55 &    1.06 & 0.98 &      0.97 &      1.00 &      1.04 &      1.03 &      1.06 &       1.06 &\textbf{1.09}&       1.07  \\
168  & 0.53 &    0.99 & 0.94 &      0.98 &      1.02 &      1.03 &      1.06 &\textbf{1.07}&       1.06 &       1.03 &\textbf{1.07} \\
189  & 0.54 &    1.07 & 0.96 &      1.05 &      1.08 &      1.08 &      1.08 &      1.08 &       1.08 &       1.09 &\textbf{1.11} \\
210  & 0.51 &    1.05 & 0.94 &      1.06 &      1.05 &\textbf{1.08}&      1.07 &      1.06 &\textbf{1.08}&       1.03 &       1.05  \\
231  & 0.54 &    0.97 & 0.92 &\textbf{0.99}&      0.97 &      0.97 &      0.98 &      0.97 &       0.97 &       0.96 &       0.96  \\
252  & 0.51 &    0.91 & 0.88 &\textbf{0.96}&      0.95 &      0.92 &      0.94 &      0.92 &       0.91 &       0.91 &       0.94  \\
\midrule 
300  & 0.52 &    0.90 & 0.91 &      0.92 &      0.88 &      0.92 &      0.92 &      0.91 &       0.93 &       0.92 &\textbf{0.95} \\
350  & 0.49 &\textbf{1.01}& 0.93 &      1.00 &      0.98 &      0.96 &      0.98 &      0.99 &       1.00 &       1.00 &       1.00  \\
400  & 0.52 &    1.01 & 0.93 &\textbf{1.03}&      1.00 &      0.99 &      0.99 &      0.99 &       1.00 &       0.98 &       1.01  \\
500  & 0.52 &    0.92 & 0.91 &\textbf{0.94}&      0.92 &      0.92 &      0.93 &      0.93 &\textbf{0.94}&       0.91 &       0.91  \\
750  & 0.49 &    0.87 & 0.90 &\textbf{0.94}&\textbf{0.94}&      0.89 &      0.89 &      0.87 &       0.88 &       0.88 &       0.89  \\
1000 & 0.51 &    0.92 &\textbf{0.95}&      0.94 &      0.92 &      0.91 &      0.91 &      0.89 &       0.89 &       0.90 &       0.91  \\
1500 & 0.47 &    0.91 & 0.94 &\textbf{0.95}&\textbf{0.95}&\textbf{0.95}&\textbf{0.95}&      0.90 &       0.94 &       0.91 &       0.90  \\
2000 & 0.43 &    0.87 & 0.88 &      0.86 &      0.90 &\textbf{0.91}&      0.86 &      0.89 &       0.88 &       0.89 &       0.89  \\\bottomrule
\end{tabular}

\caption{Sharpe-Ratio of the long-only portfolios obtained; rebalancing every 21 days. Bold entries are the optimal values for each $\tau_{in}$.}
\label{tab:SRlong}
\end{sidewaystable}

\setcounter{table}{5}
\begin{sidewaystable}
\begin{tabular}{lcccccccccc}
\toprule
& \multicolumn{2}{l}{Covariance matrix estimators } \\ \cmidrule{2-11}
$\tau_{in}$&  Sample &  CV &  $1-$BAHC &  $2-$BAHC &  $3-$BAHC &  $4-$BAHC &  $7-$BAHC &  $11-$BAHC &  $18-$BAHC &  $30-$BAHC \\\midrule
21   &     558 & 566 &       474 &       475 &       473 &       473 &       470 &        468 &        467 &\textbf{465} \\
42   &     556 & 560 &       494 &       492 &       492 &       492 &       491 &        491 &\textbf{490}&\textbf{490} \\
63   &     551 & 554 &       501 &\textbf{496}&       497 &       498 &       498 &        498 &        497 &        498  \\
84   &     546 & 549 &       502 &\textbf{495}&       497 &       498 &       498 &        498 &        498 &        499  \\
105  &     544 & 543 &       502 &\textbf{494}&       496 &       497 &       498 &        498 &        499 &        499  \\
126  &     540 & 540 &       501 &\textbf{493}&       494 &       496 &       497 &        497 &        498 &        498  \\
147  &     536 & 536 &       499 &\textbf{490}&       493 &       495 &       496 &        497 &        497 &        498  \\
168  &     533 & 532 &       498 &\textbf{489}&       491 &       493 &       495 &        496 &        496 &        497  \\
189  &     530 & 529 &       497 &\textbf{486}&       488 &       489 &       492 &        493 &        493 &        494  \\
210  &     527 & 525 &       495 &\textbf{484}&       485 &       487 &       490 &        490 &        491 &        492  \\
231  &     524 & 521 &       493 &\textbf{481}&       483 &       485 &       487 &        488 &        489 &        490  \\
252  &     521 & 518 &       491 &\textbf{479}&\textbf{479}&       482 &       484 &        486 &        487 &        487  \\
\midrule 
300  &     512 & 511 &       486 &\textbf{473}&\textbf{473}&       475 &       479 &        481 &        481 &        482  \\
350  &     504 & 503 &       480 &\textbf{467}&\textbf{467}&       468 &       472 &        474 &        475 &        476  \\
400  &     496 & 496 &       474 &       461 &\textbf{460}&       463 &       466 &        468 &        470 &        470  \\
500  &     479 & 482 &       462 &       449 &\textbf{448}&       450 &       454 &        456 &        458 &        459  \\
750  &     441 & 449 &       434 &       421 &\textbf{420}&       421 &       425 &        428 &        429 &        430  \\
1000 &     405 & 416 &       404 &       393 &\textbf{391}&       392 &       396 &        398 &        398 &        399  \\
1500 &     340 & 353 &       347 &\textbf{337}&\textbf{337}&\textbf{337}&       339 &        341 &        341 &        341  \\
2000 &     286 & 294 &       292 &       286 &       286 &\textbf{285}&       286 &        287 &        287 &        286  \\\bottomrule
\end{tabular}

\caption{$n_{90}$ of the long-short portfolios; rebalancing every 21 days. Bold entries are the optimal values for each $\tau_{in}$.}
\label{tab:N90longshort}
\end{sidewaystable}

\setcounter{table}{6}
\begin{sidewaystable}
\begin{tabular}{lcccccccccc}
\toprule
& \multicolumn{2}{l}{Covariance matrix estimators } \\ \cmidrule{2-11}
$\tau_{in}$&  Sample &    CV &  $1-$BAHC &  $2-$BAHC &  $3-$BAHC &  $4-$BAHC &  $7-$BAHC &  $11-$BAHC &  $18-$BAHC &  $30-$BAHC \\\midrule
21   &    27.5 & 155.2 &\textbf{10.3}&      14.8 &      18.8 &      21.7 &      27.3 &       31.3 &       34.5 &       36.5  \\
42   &    14.8 &  92.4 &\textbf{10.3}&      13.1 &      15.6 &      17.5 &      21.2 &       23.8 &       25.8 &       26.8  \\
63   &    13.6 &  73.9 &\textbf{10.8}&      13.0 &      14.8 &      16.3 &      19.1 &       21.2 &       22.8 &       23.5  \\
84   &    13.2 &  64.3 &\textbf{11.2}&      13.2 &      14.7 &      15.8 &      18.1 &       19.8 &       21.2 &       21.8  \\
105  &    13.5 &  56.9 &\textbf{11.7}&      13.5 &      14.9 &      15.9 &      17.9 &       19.4 &       20.5 &       21.1  \\
126  &    13.6 &  53.7 &\textbf{12.2}&      13.9 &      15.1 &      15.9 &      17.6 &       18.9 &       20.0 &       20.5  \\
147  &    13.8 &  50.5 &\textbf{12.4}&      14.0 &      15.0 &      15.8 &      17.3 &       18.4 &       19.3 &       19.8  \\
168  &    13.9 &  48.2 &\textbf{12.6}&      14.1 &      15.0 &      15.7 &      17.0 &       18.0 &       18.9 &       19.3  \\
189  &    14.0 &  46.4 &\textbf{12.7}&      14.1 &      14.9 &      15.6 &      16.7 &       17.6 &       18.4 &       18.8  \\
210  &    14.1 &  44.6 &\textbf{12.9}&      14.1 &      14.9 &      15.5 &      16.5 &       17.3 &       18.0 &       18.5  \\
231  &    14.1 &  43.5 &\textbf{13.0}&      14.1 &      14.9 &      15.4 &      16.3 &       17.0 &       17.7 &       18.1  \\
252  &    14.0 &  42.4 &\textbf{13.0}&      14.1 &      14.7 &      15.2 &      16.1 &       16.8 &       17.4 &       17.7  \\
\midrule 
300  &    14.0 &  40.0 &\textbf{13.1}&      13.9 &      14.4 &      14.8 &      15.5 &       16.1 &       16.7 &       17.0  \\
350  &    14.2 &  38.3 &\textbf{13.3}&      14.1 &      14.5 &      14.9 &      15.5 &       16.0 &       16.5 &       16.8  \\
400  &    14.2 &  36.7 &\textbf{13.4}&      14.0 &      14.4 &      14.7 &      15.2 &       15.7 &       16.1 &       16.4  \\
500  &    14.2 &  34.2 &\textbf{13.5}&      14.1 &      14.5 &      14.7 &      15.0 &       15.3 &       15.7 &       15.9  \\
750  &    14.5 &  30.3 &\textbf{14.1}&      14.5 &      14.4 &      14.5 &      14.8 &       15.0 &       15.2 &       15.5  \\
1000 &    15.1 &  28.2 &      14.8 &      14.8 &\textbf{14.6}&      14.7 &      14.9 &       15.1 &       15.4 &       15.6  \\
1500 &    15.9 &  25.5 &      15.3 &      15.3 &\textbf{15.0}&      15.2 &      15.4 &       15.6 &       15.8 &       16.0  \\
2000 &    15.4 &  22.4 &\textbf{14.4}&      14.7 &\textbf{14.4}&      14.5 &      14.8 &       14.9 &       15.1 &       15.2  \\\bottomrule
\end{tabular}

\caption{Average number of effective assets $n_{\textrm{eff}}$ of the long-only portfolios; rebalancing every 21 days. Bold entries are the optimal values for each $\tau_{in}$.}
\label{tab:Nefflong}
\end{sidewaystable}

\setcounter{table}{7}
\begin{sidewaystable}
\begin{tabular}{lcccccccccc}
\toprule
& \multicolumn{2}{l}{Covariance matrix estimators } \\ \cmidrule{2-11}
$\tau_{in}$&  Sample &   CV &  $1-$BAHC &  $2-$BAHC &  $3-$BAHC &  $4-$BAHC &  $7-$BAHC &  $11-$BAHC &  $18-$BAHC &  $30-$BAHC \\\midrule
21   &    2.53 &\textbf{1.65}&      2.43 &      2.45 &      2.34 &      2.26 &      2.12 &       2.03 &       1.95 &       1.91  \\
42   &    2.50 &\textbf{1.97}&      2.69 &      2.81 &      2.71 &      2.64 &      2.51 &       2.42 &       2.36 &       2.33  \\
63   &    2.44 &\textbf{2.17}&      2.85 &      3.04 &      2.96 &      2.89 &      2.77 &       2.69 &       2.63 &       2.61  \\
84   &    2.70 &\textbf{2.33}&      2.96 &      3.20 &      3.13 &      3.07 &      2.95 &       2.88 &       2.83 &       2.82  \\
105  &    2.98 &\textbf{2.47}&      3.04 &      3.32 &      3.28 &      3.22 &      3.11 &       3.04 &       3.00 &       3.00  \\
126  &    3.25 &\textbf{2.55}&      3.11 &      3.42 &      3.39 &      3.34 &      3.24 &       3.17 &       3.13 &       3.14  \\
147  &    3.52 &\textbf{2.64}&      3.16 &      3.51 &      3.51 &      3.46 &      3.36 &       3.30 &       3.26 &       3.27  \\
168  &    3.78 &\textbf{2.72}&      3.21 &      3.58 &      3.59 &      3.56 &      3.46 &       3.40 &       3.37 &       3.39  \\
189  &    4.04 &\textbf{2.79}&      3.24 &      3.63 &      3.66 &      3.63 &      3.55 &       3.49 &       3.46 &       3.48  \\
210  &    4.31 &\textbf{2.87}&      3.27 &      3.68 &      3.73 &      3.70 &      3.62 &       3.57 &       3.54 &       3.56  \\
231  &    4.59 &\textbf{2.92}&      3.29 &      3.72 &      3.78 &      3.76 &      3.69 &       3.64 &       3.61 &       3.64  \\
252  &    4.87 &\textbf{2.98}&      3.31 &      3.76 &      3.83 &      3.82 &      3.75 &       3.70 &       3.68 &       3.71  \\
\midrule 
300  &  498.38 &\textbf{3.11}&      3.34 &      3.83 &      3.93 &      3.93 &      3.87 &       3.83 &       3.81 &       3.85  \\
350  &  327.54 &\textbf{3.21}&      3.38 &      3.88 &      4.01 &      4.02 &      3.98 &       3.94 &       3.93 &       3.97  \\
400  &  740.75 &\textbf{3.31}&      3.40 &      3.93 &      4.08 &      4.10 &      4.07 &       4.03 &       4.03 &       4.07  \\
500  &  380.61 & 3.47 &\textbf{3.43}&      4.00 &      4.19 &      4.22 &      4.22 &       4.19 &       4.19 &       4.24  \\
750  &  207.73 & 3.75 &\textbf{3.48}&      4.11 &      4.36 &      4.42 &      4.45 &       4.45 &       4.46 &       4.51  \\
1000 &   17.89 & 3.91 &\textbf{3.50}&      4.16 &      4.44 &      4.51 &      4.56 &       4.56 &       4.58 &       4.64  \\
1500 &    7.10 & 3.97 &\textbf{3.47}&      4.08 &      4.36 &      4.44 &      4.52 &       4.54 &       4.56 &       4.62  \\
2000 &    5.69 & 3.82 &\textbf{3.36}&      3.88 &      4.12 &      4.21 &      4.31 &       4.33 &       4.35 &       4.40  \\\bottomrule
\end{tabular}

\caption{Gross leverage of the long-short portfolios; rebalancing every 21 days. Bold entries are the optimal values for each $\tau_{in}$.}\label{tab:Grosslongshort}
\end{sidewaystable}

\setcounter{table}{8}
\begin{sidewaystable}
\begin{tabular}{lcccccccccc}
\toprule
& \multicolumn{2}{l}{Covariance matrix estimators } \\ \cmidrule{2-11}
$\tau_{in}$&  Sample &   CV &  $1-$BAHC &  $2-$BAHC &  $3-$BAHC &  $4-$BAHC &  $7-$BAHC &  $11-$BAHC &  $18-$BAHC &  $30-$BAHC \\\midrule
21   &    3.75 &\textbf{1.54}&      2.93 &      3.03 &      2.89 &      2.79 &      2.61 &       2.50 &       2.41 &       2.36  \\
42   &    3.01 &\textbf{1.36}&      2.27 &      2.58 &      2.52 &      2.47 &      2.35 &       2.27 &       2.22 &       2.19  \\
63   &    2.24 &\textbf{1.25}&      1.81 &      2.23 &      2.23 &      2.20 &      2.12 &       2.07 &       2.03 &       2.02  \\
84   &    2.24 &\textbf{1.18}&      1.59 &      2.05 &      2.09 &      2.07 &      2.02 &       1.97 &       1.95 &       1.94  \\
105  &    2.35 &\textbf{1.19}&      1.40 &      1.89 &      1.96 &      1.96 &      1.92 &       1.88 &       1.86 &       1.87  \\
126  &    2.45 &\textbf{1.06}&      1.25 &      1.76 &      1.85 &      1.86 &      1.84 &       1.81 &       1.79 &       1.80  \\
147  &    2.62 &\textbf{1.02}&      1.15 &      1.67 &      1.78 &      1.79 &      1.78 &       1.75 &       1.73 &       1.74  \\
168  &    2.79 &\textbf{0.98}&      1.05 &      1.59 &      1.71 &      1.73 &      1.72 &       1.69 &       1.68 &       1.69  \\
189  &    2.93 &\textbf{0.96}&\textbf{0.96}&      1.50 &      1.64 &      1.68 &      1.67 &       1.65 &       1.64 &       1.65  \\
210  &    3.10 & 0.97 &\textbf{0.91}&      1.46 &      1.60 &      1.64 &      1.64 &       1.62 &       1.61 &       1.62  \\
231  &    3.28 & 0.92 &\textbf{0.86}&      1.41 &      1.56 &      1.61 &      1.61 &       1.59 &       1.57 &       1.58  \\
252  &    3.46 & 0.89 &\textbf{0.80}&      1.36 &      1.52 &      1.56 &      1.57 &       1.56 &       1.54 &       1.55  \\
\midrule 
300  &  916.40 & 0.86 &\textbf{0.73}&      1.29 &      1.46 &      1.51 &      1.53 &       1.51 &       1.50 &       1.51  \\
350  &  567.25 & 0.82 &\textbf{0.66}&      1.23 &      1.41 &      1.46 &      1.49 &       1.47 &       1.45 &       1.46  \\
400  & 1374.68 & 0.80 &\textbf{0.61}&      1.18 &      1.37 &      1.43 &      1.45 &       1.44 &       1.42 &       1.42  \\
500  &  664.68 & 0.74 &\textbf{0.53}&      1.10 &      1.30 &      1.36 &      1.39 &       1.37 &       1.35 &       1.35  \\
750  &  346.99 & 0.66 &\textbf{0.41}&      0.98 &      1.19 &      1.26 &      1.30 &       1.28 &       1.26 &       1.25  \\
1000 &   13.92 & 0.60 &\textbf{0.35}&      0.89 &      1.11 &      1.18 &      1.22 &       1.20 &       1.18 &       1.17  \\
1500 &    1.53 & 0.49 &\textbf{0.28}&      0.77 &      0.96 &      1.03 &      1.07 &       1.05 &       1.02 &       1.01  \\
2000 &    0.88 & 0.40 &\textbf{0.24}&      0.64 &      0.82 &      0.88 &      0.91 &       0.90 &       0.88 &       0.86  \\\bottomrule
\end{tabular}

\caption{$\gamma$ of the long-short portfolios; rebalancing every 21 days. Bold entries are the optimal values for each $\tau_{in}$.}\label{tab:Gammalongshort}
\end{sidewaystable}

\setcounter{table}{9}
\begin{sidewaystable}
\begin{tabular}{lcccccccccc}
\toprule
& \multicolumn{2}{l}{Covariance matrix estimators } \\ \cmidrule{2-11}
$\tau_{in}$&  Sample &   CV &  $1-$BAHC &  $2-$BAHC &  $3-$BAHC &  $4-$BAHC &  $7-$BAHC &  $11-$BAHC &  $18-$BAHC &  $30-$BAHC \\\midrule
21   &    1.61 &\textbf{1.37}&      1.51 &      1.53 &      1.53 &      1.53 &      1.53 &       1.53 &       1.53 &       1.53  \\
42   &    1.47 &\textbf{1.04}&      1.13 &      1.14 &      1.14 &      1.14 &      1.14 &       1.14 &       1.15 &       1.15  \\
63   &    1.19 &\textbf{0.82}&      0.87 &      0.89 &      0.89 &      0.89 &      0.89 &       0.89 &       0.90 &       0.91  \\
84   &    1.00 &\textbf{0.69}&      0.76 &      0.77 &      0.77 &      0.78 &      0.78 &       0.78 &       0.79 &       0.79  \\
105  &    0.85 &\textbf{0.60}&      0.66 &      0.66 &      0.67 &      0.67 &      0.67 &       0.68 &       0.68 &       0.69  \\
126  &    0.76 &\textbf{0.52}&      0.58 &      0.59 &      0.60 &      0.60 &      0.61 &       0.61 &       0.61 &       0.62  \\
147  &    0.66 &\textbf{0.46}&      0.51 &      0.52 &      0.53 &      0.54 &      0.54 &       0.54 &       0.54 &       0.55  \\
168  &    0.59 &\textbf{0.42}&      0.45 &      0.46 &      0.47 &      0.48 &      0.48 &       0.48 &       0.48 &       0.49  \\
189  &    0.55 &\textbf{0.38}&      0.41 &      0.43 &      0.44 &      0.44 &      0.44 &       0.45 &       0.45 &       0.45  \\
210  &    0.51 &\textbf{0.36}&      0.39 &      0.40 &      0.42 &      0.42 &      0.43 &       0.43 &       0.43 &       0.43  \\
231  &    0.47 &\textbf{0.34}&      0.36 &      0.38 &      0.39 &      0.39 &      0.39 &       0.39 &       0.40 &       0.40  \\
252  &    0.43 &\textbf{0.31}&      0.33 &      0.35 &      0.36 &      0.36 &      0.37 &       0.37 &       0.37 &       0.37  \\
\midrule 
300  &    0.38 &\textbf{0.28}&      0.30 &      0.32 &      0.33 &      0.33 &      0.33 &       0.33 &       0.33 &       0.34  \\
350  &    0.34 &\textbf{0.25}&      0.27 &      0.29 &      0.30 &      0.30 &      0.30 &       0.30 &       0.30 &       0.31  \\
400  &    0.30 &\textbf{0.23}&      0.24 &      0.26 &      0.28 &      0.28 &      0.28 &       0.28 &       0.28 &       0.28  \\
500  &    0.25 &\textbf{0.20}&      0.21 &      0.23 &      0.24 &      0.24 &      0.24 &       0.24 &       0.24 &       0.24  \\
750  &    0.19 &\textbf{0.15}&      0.16 &      0.19 &      0.19 &      0.20 &      0.20 &       0.20 &       0.20 &       0.20  \\
1000 &    0.16 &\textbf{0.13}&      0.14 &      0.16 &      0.17 &      0.17 &      0.17 &       0.17 &       0.17 &       0.17  \\
1500 &    0.11 &\textbf{0.09}&      0.10 &      0.13 &      0.13 &      0.13 &      0.13 &       0.13 &       0.13 &       0.13  \\
2000 &    0.09 &\textbf{0.08}&      0.09 &      0.11 &      0.12 &      0.12 &      0.12 &       0.11 &       0.11 &       0.11  \\\bottomrule
\end{tabular}

\caption{$\gamma$ of the long-only portfolios; rebalancing every 21 days. Bold entries are the optimal values for each $\tau_{in}$.}
\label{tab:Gammalong}
\end{sidewaystable}

\end{document}